\documentclass[%
aps,
prd,
twocolumn,
superscriptaddress,
preprintnumbers,
nofootinbib,
amsmath,amssymb,
]{revtex4-2}

\usepackage{style}
\usepackage{aas_macros}

\def\be{\begin{equation}}
\def\ee{\end{equation}}
\def\ba{\begin{aligned}}
\def\ea{\end{aligned}}
\def\bi{\begin{itemize}}
\def\ei{\end{itemize}}
\def\ben{\begin{enumerate}}
\def\een{\end{enumerate}}

\newcommand{\GWB}{\scriptscriptstyle\rm GWB}
\newcommand{\BF}{\scriptscriptstyle\rm BF}

\newcommand{\LR}{\scriptscriptstyle\rm LR}

\newcommand{\D}{{\rm d}}

\newcommand{\bh}{\vec{h}}
\newcommand{\bx}{\vec{x}}
\newcommand{\bC}{{C}}

\newcommand{\btheta}{{\theta}}
\newcommand{\bdt}{{\delta t}}
\newcommand{\brho}{{\rho}}

\begin{document}

\preprint{Imperial--TP--2026--AM--04}

\title{Opening the Black Box: What Neural Networks Learn from\\ Pulsar Timing Array Data}

\author{James Alvey\,\orcidlink{0000-0003-2020-0803}}
\affiliation{Kavli Institute for Cosmology Cambridge, Madingley Road, Cambridge CB3 0HA, United Kingdom}
\affiliation{Institute of Astronomy, University of Cambridge, Madingley Road, Cambridge CB3 0HA, United Kingdom}

\author{Andrea Mitridate\,\orcidlink{0000-0003-2898-5844}}
\affiliation{Abdus Salam Centre for Theoretical Physics, Imperial College, London, SW7 2AZ, United Kingdom}

\author{Mauro Pieroni\,\orcidlink{0000-0003-0665-266X}}
\affiliation{
Instituto de Estructura de la Materia (IEM), CSIC, Serrano 121, 28006 Madrid, Spain}

\author{Joseph D.~Romano\,\orcidlink{0000-0003-4915-3246}}
\affiliation{Department of Physics and Astronomy, University of Texas Rio Grande Valley, Brownsville TX 78520, USA}

\begin{abstract}
\noindent In recent years, simulation-based inference (SBI) methods have been proposed to address several data-analysis challenges faced by existing and planned gravitational-wave experiments. For example, SBI classification has recently been shown to significantly improve the prospects for detecting anisotropies in pulsar timing array (PTA) data. In this work, we use a simple toy model of a PTA to provide a more pedagogical explanation of how, and under which circumstances, SBI-based detection can improve on classical detection statistics for gravitational-wave background anisotropies.
\end{abstract}

\maketitle
\tableofcontents

\section{Introduction}
\label{sec:intro}
\noindent Simulation-based inference (SBI) has been proposed to address several challenges in gravitational-wave (GW) data analysis (see, e.g., Ref.~\cite{Cranmer:2019eaq} for a review and Refs.~\cite{Shih:2023jme, Lai:2025xov, Laal:2024trp, Lemke:2026gxj, Dandapat:2026kfd} for applications in the PTA context). Because SBI requires only the ability to simulate data, and not to evaluate a likelihood, it can be applied when the likelihood is unknown or intractable, or when sampling from it is prohibitively slow.

Over the past few years, pulsar timing array (PTA) collaborations have reported mounting evidence for a nanohertz gravitational-wave background (GWB), whose characteristic signature is the Hellings-Downs pattern of correlations between the timing residuals of pairs of pulsars~\cite{Hellings:1983fr, NANOGrav:2023gor, EPTA:2023fyk, Reardon:2023gzh, Xu:2023wog, Miles:2024seg, InternationalPulsarTimingArray:2023mzf}. In pulsar timing array analyses, the likelihood is usually known. Therefore, standard Bayesian methods that sample it using Monte Carlo (MC) techniques are available (see, e.g., Ref.~\cite{enterprise}). However, despite several recent developments (see, for example, Refs.~\cite{Vallisneri:2024xfk, discovery, Gundersen:2026ows}), such analyses can be computationally prohibitive: the analysis of a single dataset may require days of CPU time~\cite{Gundersen:2026ows}, so studies demanding the analysis of many mock datasets (e.g., for deriving null distributions, or in forecast studies) can quickly become impractical. To address this computational challenge, frequentist methods have been developed as a faster alternative (see, for example, Refs.~\cite{Anholm:2008wy, Chamberlin:2014ria, Pol:2022sjn, vanHaasteren:2025siu,Hazboun:2023tiq}). However, the speed of frequentist methods often comes at the price of several assumptions that can limit their sensitivity. For example, the optimal statistic is derived in the weak-signal limit, and becomes suboptimal as the GWB signal grows relative to the noise~\cite{Anholm:2008wy, Chamberlin:2014ria}.

SBI methods can combine the strengths of both approaches: given a sufficiently flexible network, they can match the sensitivity of a full Bayesian analysis while retaining, after appropriate training, the speed of frequentist methods. 
Reference~\cite{Lemke:2026gxj} recently applied this idea to the search for anisotropy in the gravitational-wave background observed in PTA data, constructing a detection statistic that avoids several of the approximations underlying conventional frequentist searches---most notably, the assumption that the cross-correlation estimators are Gaussian distributed---and achieves substantially higher detection probabilities.

In this work, we take a closer look at the results of Ref.~\cite{Lemke:2026gxj} to better understand the origin of this improvement and when it can be expected. We do so by introducing a toy model of a two-dimensional PTA, simple enough that the Bayes factor -- the optimal detection statistic in the Neyman-Pearson sense~\cite{Neyman:1933wgr} -- can be evaluated exactly. This lets us benchmark SBI-based detection statistics not only against their classical frequentist counterparts, but against the optimal sensitivity a full Bayesian analysis would achieve. This exercise aims to help the GW community build confidence in the machine-learning methods now being developed for GW data analysis, by clarifying what these methods learn and under which conditions they can be expected to improve over classical methods.

The paper is structured as follows. In Sec.~\ref{sec:flatland} we introduce the two-dimensional toy model and derive its overlap reduction function, showing that the array is sensitive to only the monopole and quadrupole components of the GWB sky. Section~\ref{sec:likelihood} constructs the corresponding likelihood and identifies the pulsar correlations as a sufficient statistic. In Sec.~\ref{sec:det_stats}, we review the detection statistics that we compare: the Bayes factor and its weak-signal limit, the sky-map-based frequentist statistic, and the linear and non-linear neural-network classifiers. Section~\ref{sec:results} compares these detection statistics and interprets what the networks learned from the data. We conclude in Sec.~\ref{sec:conclusions}.

%%%%%%%%%%%%%%%%%%%%%%%%%
\section{\texorpdfstring{PTA\MakeLowercase{s}}{PTAs} in Flatland}
\label{sec:flatland}

\noindent To make some of the analytical calculations and numerical comparisons presented in this work more tractable, we consider a two-dimensional (2D) toy model of a PTA. In this 2D world, the Earth (or the solar-system barycenter, to be more precise) lies at the origin of the $xy$-plane and all pulsars lie on a unit circle centered on the origin, such that the vector pointing from the Earth to the $a^{\rm th}$ pulsar in the array is given by:
\begin{equation}
\hat p_a=\cos\phi_a\,\hat x+\sin\phi_a\,\hat y\,,
\qquad a=1,2,\cdots, N_p\,,
\end{equation}
and is thus uniquely specified by an angular coordinate $\phi_a\in[0,\,2\pi)$.

%%%%%%%%%%%%%%%%%%%%%%%%%%%%%%%%%%
\subsection{Metric perturbations and two-point functions}

\noindent For this 2D example, a generic metric perturbation is given by a vector $\bh(t,\,\bx)$, which can be decomposed as a superposition of plane waves,
\begin{equation}
\label{eq:met_pert}
    \bh(t,\,\bx) = \int_{-\infty}^\infty \D f\int_0^{2\pi}\D{\phi}\,\tilde h(f,\,\phi) 
    \hat{\epsilon}(\phi)
    e^{2\pi if(t-\hat{\Omega}\cdot\bx)}\,,
\end{equation}
where $f$ is the GW frequency, $\tilde h(f,\phi)$ is a complex function satisfying $\tilde h^*(f,\phi)=\tilde h(-f,\phi)$, 
and $\hat\Omega$ is the direction of propagation of the plane wave, which can be written in polar coordinates as 
\begin{equation}
    \hat\Omega(\phi) = -\cos\phi\,\hat x-\sin\phi\,\hat y\,.
\end{equation}
Similarly, the GW polarization vector can be written in polar coordinates as
\begin{equation}
\hat\epsilon(\phi) = -\sin\phi\,\hat x +\cos\phi\,\hat y\,.
\label{e:epsilon_hat}
\end{equation}

A Gaussian and stationary GWB is completely described in terms of the one- and two-point functions of the Fourier modes, $
\tilde{h}(f,\phi)$. 
We will assume that the one-point function vanishes, i.e., $\langle \tilde{h}(f,\phi)\rangle = 0$, where $\langle \cdot \rangle$ denotes an average over the Gaussian ensemble. 
We parametrize the two-point function as
\begin{equation}
    \langle \tilde h(f,\phi)\tilde h^*(f',\phi')\rangle = \frac{1}{2}H(f,\phi)\,\delta(f-f')\delta(\phi-\phi')\,,
\end{equation}
and we will assume that the function $H(f,\phi)$ can be factorized as $H(f,\phi)\equiv H(f) P(\phi)$ with $H(f)$ being the (one-sided) GWB power spectrum and $P(\phi)$ the sky map describing the distribution of GWB power on the sky.\footnote{This factorization amounts to the assumption that the anisotropies are frequency-independent.}
The sky map is normalized such that $\int_0^{2\pi}\D \phi\>P(\phi) = 1$. 

%%%%%%%%%%%%%%%%%%%%%%%%%%%%%%%%%%
\subsection{Timing-residual response and expected correlation functions}
\noindent Assuming (in this 2D world, as in the real one) that the pulsar distances $L_a$ satisfy $f L_a \gg 1$, we can neglect the pulsar term when computing the stochastic signal produced by the GWB and retain only the Earth-term timing residuals. Given the metric perturbations from  Eq.~\eqref{eq:met_pert}, these timing residuals are:
\begin{equation}
\label{eq:res_time}
    \delta t_a(t)
    \equiv\hat{p}_a\!\cdot\!\bh(t,0)
    =\int_{-\infty}^{\infty} \D f\!\int_0^{2\pi} \D \phi\,\tilde{h}(f,\phi) F_a(\phi)e^{2\pi ift}\,,
\end{equation}
where we have defined the response function
\begin{equation}
    F_a(\phi)\equiv \hat p_a\cdot \hat\epsilon(\phi) = -\sin(\phi-\phi_a)\,.
    \label{e:dipole_response}
\end{equation}
In Fourier space, Eq.~\eqref{eq:res_time} takes the form
\begin{equation}
    \tilde{\delta t}_a(f) = \int_0^{2\pi}\D\phi\, \tilde h(f,\phi)\,F_a(\phi)\,.
\end{equation}
Therefore, we can write the two-point function for the stochastic signal produced by the GWB as:
\begin{equation}
    \langle \tilde{\delta t}_a(f)\tilde{\delta t}^*_b(f')\rangle
    =\frac{1}{2}\delta(f-f')H(f)\Gamma_{ab}\,,
\end{equation}
where 
\begin{equation}
    \Gamma_{ab} \equiv \int_0^{2\pi}\D\phi \>F_a(\phi)F_b(\phi)P(\phi)
    \label{e:Gamma_ab_def}
\end{equation}
is the (unnormalized) {\it overlap reduction function} (ORF), which in general depends on the distribution of GW power on the sky.

For an isotropic GWB (given by $P(\phi)=1/2\pi$), we obtain the 2D-analog of the Hellings-Downs (HD) correlation function~\cite{Hellings:1983fr}:
\begin{equation}
    \Gamma^{\rm iso}_{ab} \equiv \frac{1}{2\pi}\int_{0}^{2\pi}
    \D\phi\>
    F_a(\phi) F_b(\phi)
    =\frac{1}{2}\cos(\gamma_{ab})\,,
\end{equation}
where $\gamma_{ab} \equiv \phi_a-\phi_b$.
For our simple 2D toy model, the cosine dependence is expected for the pure dipolar response in Eq.~\eqref{e:dipole_response}. 

For anisotropic GWBs, we can decompose the sky-map in Fourier modes as:
\begin{equation}
    P(\phi)=\sum_{m=-\infty}^\infty P_m e^{im\phi}\,,
\label{e:P_fourier}
\end{equation}
where the reality and normalization of $P(\phi)$ imply that the Fourier coefficients satisfy $P_{-m}=P_m^*$ and $P_0=1/2\pi$. Similarly, if we define
\begin{equation}\label{eq:R_response}
    R_{ab}(\phi)
    \equiv F_a(\phi)F_b(\phi)
    =\sum_{m=-\infty}^\infty R_{ab,m} e^{im\phi}\,,
\end{equation}
then
\begin{equation} \Gamma_{ab}=2\pi\sum_m R^*_{ab,m}P_m\,.
\end{equation}
Since $\int_0^{2\pi} d\phi\, e^{im\phi}=0$ for any integer $m\neq 0$, it follows that $R_{ab,m}$ takes the form
\begin{equation}  R_{ab,m}=\Gamma^{\rm iso}_{ab}\,\delta_{0m}-\frac{1}{4}\left[e^{-i(\phi_a+\phi_b)}\delta_{2m}+e^{i(\phi_a+\phi_b)}\delta_{-2m}\right]\,,
\end{equation}
from which we obtain the ORF for our 2D PTA:
\begin{align}
&\Gamma_{ab}=\Gamma^{\rm iso}_{ab}+\Gamma_{ab}^{\rm aniso}\,,
\quad{\rm where}
\label{e:Gamma_ab_final}\\
&\Gamma_{ab}^{\rm iso}
\equiv\frac{1}{2}\,\cos(\phi_a-\phi_b)\,,
\label{e:Gamma_iso}\\
&\Gamma_{ab}^{\rm aniso}
\equiv -\pi\left(P_2^{\rm r} \cos(\phi_a+\phi_b)-P_2^{\rm i}\sin(\phi_a+\phi_b)\right).
\label{e:Gamma_aniso}
\end{align}
In the above expression for $\Gamma_{ab}^{\rm aniso}$, we have defined $P_2^{\rm r}$ and $P_2^{\rm i}$ as the real and imaginary parts of $P_2$, respectively.  Therefore, a 2D PTA is sensitive to only three GWB sky parameters: the monopole $P_0$ (fixed at $P_0=1/2\pi$) and the quadrupole components $P_2^{\rm r}$ and $P_2^{\rm i}$.

%%%%%%%%%%%%%%%%%%%%%%%%%%
\section{PTA likelihood}
\label{sec:likelihood}

\noindent Here, we complete our specification of our toy model PTA by including the effects of pulsar noise and the prior probability distributions for the model parameters.

%%%%%%%%%%%%%%%%%%%%%%%%%%%%%%%%%%
\subsection{Cross-power spectra}

\noindent  In addition to the GWB signal discussed in the previous section, stochastic sources other than GW signals will contribute to the timing residuals. Typically, these noise sources are modeled as additional Gaussian processes, such that the two-point function of these timing residuals reads
\begin{equation}
    \langle\tilde{\delta t}_a(f)\tilde{\delta t}^*_b(f')\rangle=\frac{1}{2}\delta(f-f')S_{ab}(f)\,,
\label{e:two-point}
\end{equation}
where
\begin{equation}
S_{ab}(f)\equiv \delta_{ab}N_a(f)
+\Gamma_{ab}H(f)
\label{e:P_ab(f)}
\end{equation}
is the cross-power spectral density for pulsars $a$ and $b$.
Here, $N_a(f)$ is the (one-sided) power spectrum of the noise for pulsar $a$.
The $\delta_{ab}$ corresponds to the assumption that the pulsar noise is uncorrelated across pulsars in the array.

For a PTA operating for a finite observing time, $T$, it is customary to write the timing residuals as a discrete Fourier sum:
\begin{equation}
    \delta t_a(t)=\sum_j\tilde{\delta t}_{aj}e^{2\pi if_j t}\,,
\end{equation}
where the frequency components are $f_j\equiv j/T$, and the sum runs over 
$j\in-N_f, \ldots,-1,1,\ldots,N_f$. Multiplying the above equation by $e^{-2\pi i f_j t}/T$ and integrating over $t\in[-T/2,T/2]$, we find that the values of the discrete Fourier components, $\tilde{\delta t}_{aj}$, are related to $\tilde{\delta t}_a(f)$ by:
\begin{equation}
    \tilde{\delta t}_{aj}=\int_{-\infty}^{\infty}\!\!\D f\,\tilde{\delta t}_a(f)\,{\rm sinc}\big[\pi(f-f_j)T\big]\,.
    \label{e:Fseries_exp}
\end{equation}
If we define the cross-power matrix $S_{aj, bk}$ via
\begin{equation}
\langle \tilde{\delta t}_{aj}\tilde{\delta t}^*_{bk}\rangle
\equiv \frac{1}{2} S_{aj, bk}
\label{e:Pmatrix_def}
\end{equation}
(cf. Eq.~\eqref{e:two-point}), then we can show that
\begin{equation}
    S_{aj,bk}=\int_{-\infty}^{\infty}\!\!\D f\,S_{ab}(f){\rm sinc}\big[\pi(f-f_j)T\big]{\rm sinc}\big[\pi(f-f_k)T\big]\,.
    \label{e:Pmatrix}
\end{equation}
Equations~\eqref{e:Fseries_exp}-\eqref{e:Pmatrix} show that: (i) the $j^{\rm th}$ Fourier coefficient is, in general, not simply equal to the Fourier transform of the data evaluated at the bin center $f_j$, and (ii) different Fourier coefficients are, in general, correlated with one another. 

%%%%%%%%%%%%%%%%%%%%%%%%%%%%%%%%%%
\subsection{Likelihood function and covariance matrix}

\noindent The likelihood function for the timing residuals is given by\footnote{Standard PTA analyses define the likelihood in the time domain, since real TOAs are irregularly sampled. Because the TOAs in our toy model are equally spaced, we can instead work directly in the frequency domain. We also neglect the marginalization over the timing model.}
\begin{equation}
    p(\bdt|\btheta) = \frac{\exp{
    \Big(-\frac{1}{2}\tilde{\bdt}^\dagger}\cdot\bC^{-1}(\btheta)\cdot\tilde{\bdt}\Big)}{\sqrt{{\rm det}(2\pi \bC(\btheta))}}\,,
    \label{e:likelihood}    
\end{equation}
where the dot products in the argument of the exponential are defined by
\begin{equation}
\begin{aligned}    \tilde{\bdt}^\dagger\!\cdot\!\bC^{-1}(\btheta)\!\cdot\!\tilde{\bdt}=
\sum_{a,b=1}^{N_ p}\sum_{j,k=-N_f}^{N_f}\tilde{\delta t}{}^*_{aj}\,[C^{-1}(\btheta)]_{aj,bk}\,\tilde{\delta t}_{bk}\,.
\label{e:arg_exp}
\end{aligned}
\end{equation}
Here $\tilde{\bdt}\equiv \{\tilde{\delta t}_{aj}\}$ is a vector of length $N_p N_s$ containing the Fourier components for all pulsars in the array, and 
$\bC(\btheta)\equiv \{C(\btheta)_{aj,bk}\}$ is a $(N_p N_s)\times(N_p N_s)$ matrix that depends on the model parameters $\btheta$ for either the isotropic or anisotropic signal models (see Eq.~\eqref{e:params0} and Eq.~\eqref{e:params1} for an explicit list of parameters for our toy model hypotheses).

The covariance matrix $\bC(\btheta)$ is related to the cross-power matrix defined in Eq.~\eqref{e:Pmatrix_def} via 
\begin{equation}
C_{aj,bk}=\frac{1}{2}S_{aj,bk}\,.
\end{equation}
The $\btheta$ dependence of $\bC(\btheta)$ arises via the anisotropy parameters $P_2^{\rm r}$ and $P_2^{\rm i}$ that enter $\Gamma_{ab}$ (see Eq.~\eqref{e:Gamma_ab_final}) and the power spectral densities $H(f)$ and $N_a(f)$ that enter $\bC(\btheta)$ via $S_{ab}(f)$ (see Eq.~\eqref{e:P_ab(f)}).

To further simplify our analysis, we additionally assume that the noise and GWB are described by band-limited white power spectra for $-f_{\rm max}<f<f_{\rm max}$, and that they vanish for $|f|>f_{\rm max}$.
We choose our sampling period $\Delta t$ such that the Nyquist frequency $f_{\rm Nyq}\equiv 1/2\Delta t\ge f_{\rm max}$, and we assume that the number of time samples $N_s\equiv T/\Delta t\gg 1$.
Under these simplifying assumptions, the covariance matrix for the Fourier series coefficients is approximately diagonal:
\begin{equation}
    C_{aj,bk}\approx \frac{1}{N_s} \delta_{jk}\left(\delta_{ab}\sigma_n^2 + \Gamma_{ab}\sigma_{\GWB}^2\right)\,,
\label{e:C_aj,bk}
\end{equation}
where $\sigma^2_{\GWB}$ is the variance of the white GWB signal, and $\sigma_n^2$ is the variance of the white pulsar noise, assumed to be the same for all pulsars.
Recall that $N_s=2N_f$, where $N_f$ is the number of positive frequency bins.

Note that the covariance matrix $C_{aj,bk}$ given in Eq.~\eqref{e:C_aj,bk} has the form of a Kronecker product of a smaller $N_p\times N_p$ matrix 
\begin{equation}
C_{ab}\equiv \frac{1}{N_s} (\delta_{ab}\,\sigma^2_n+ \Gamma_{ab}\,\sigma^2_{\GWB})
\label{e:C_ab}
\end{equation}
and the $N_s\times N_s$ identity matrix $\delta_{jk}$.
Thus, the inverse $(C^{-1})_{aj, bk}$ also has a Kronecker product form, involving $(C^{-1})_{ab}$ and $\delta_{jk}$.
This means that the argument of the exponential (see Eq.~\eqref{e:arg_exp}) of the Gaussian likelihood function can be simplified to
\be
-\frac{1}{2}
\tilde{\bdt}^\dagger\!\cdot\!\bC^{-1}(\btheta)\!\cdot\!\tilde{\bdt}
= -\frac{1}{2}{\rm Tr}(\bC^{-1}(\btheta)\,\brho)\,,
\ee
where 
\begin{equation}\label{e:rho_ab}
    \rho_{ab}\equiv \sum_{j=-N_f}^{N_f} \tilde{\delta t}_{aj}\tilde{\delta t}{}^*_{bj}
\end{equation}
is a sufficient statistic for the analysis.
The existence of this sufficient statistic allows us to compress the data by a factor of $N_s=2N_f$, which is the number of independent samples of white noise and white GWB.

%%%%%%%%%%%%%%%%%%%%%%%%%%%%%%%%%%
\subsection{Model parameters and model evidences}

\noindent Throughout, we denote by $\bC_0\equiv \bC(\btheta_0)$ and $\bC_1\equiv \bC(\btheta_1)$ the covariance matrices for the null and signal hypotheses, $H_0$ and $H_1$, parametrized by
\begin{align}
    \btheta_0 &\equiv \{\sigma^2_n,\,\sigma^2_{\GWB}\}\,,
    \label{e:params0}
    \\
    \btheta_1 &\equiv \{\sigma^2_n,\,\sigma^2_{\GWB},\,P_2^{\rm r},\,P_2^{\rm i}\}\,.
    \label{e:params1}
\end{align}
When performing Bayesian inference, we also need to specify prior probability distributions $\pi(\btheta_0|H_0)$ and $\pi(\btheta_1|H_1)$ for the model parameters. The model evidence, or marginalized likelihood, is then obtained by integrating the likelihood against the prior over the model parameters:
\begin{equation}
    p_0(\bdt)\equiv p(\bdt|H_0)
    \equiv \int \D \btheta_0\>
    p(\bdt|\btheta_0, H_0)
    \pi(\btheta_0|H_0)\,,
    \label{e:p0}
\end{equation}
and similarly for $p_1(\bdt)\equiv p(\bdt|H_1)$.

%%%%%%%%%%%%%%%%%%%%%%%%%%%%%%%%%%
\section{Detection statistics}
\label{sec:det_stats}

\noindent In this section, we introduce a range of anisotropy detection statistics commonly used in the literature (see e.g. Ref.~\cite{Romano:2016dpx} for a review in the context of stochastic searches) to distinguish between the null hypothesis, $H_0$, and the signal hypothesis, $H_1$, and discuss their expected relative performance.

\subsection{Bayes factor}
\label{sec:BF_det_stats}

\noindent A common detection statistic is the log of the Bayes factor between the two hypotheses, defined as the ratio of their evidences:
\begin{equation}
\Lambda_{\BF}(\bdt)\equiv \ln{\mathcal B}_{10}(\bdt)\,,
\quad
{\mathcal B}_{10}(\bdt)\equiv \frac{p(\bdt|H_1)}{p(\bdt|H_0)}\,.
\label{e:lnBF}
\end{equation} 
It can be shown~\cite{Fowlie:2021zyf,Neyman:1933wgr} that the Bayes factor satisfies the Neyman-Pearson criterion, i.e., it is the optimal statistic in the sense that it maximizes the prior-averaged detection probability (DP) for fixed false-alarm probability (FAP).

For \emph{simple hypotheses}, which have no free parameters, the log of the Bayes factor reduces to the log of the likelihood ratio:
\begin{equation}
\Lambda_{\BF}(\bdt)\rightarrow
\Lambda_{\LR}(\bdt)
\equiv \ln\left(\frac{p(\bdt|\bar\btheta_1,H_1)}{p(\bdt|\bar\btheta_0,H_0)}\right)\,,
\label{e:LR}
\end{equation}
where $\bar\btheta_0$ and $\bar\theta_1$ denote \emph{fixed} values of the parameters $\btheta_0$ and $\btheta_1$ associated with $H_0$ and $H_1$.
Furthermore, if the likelihood functions are defined by zero-mean Gaussian-distributed data, then the log Bayes factor reads:
\begin{equation}
\Lambda_{\LR}(\bdt)
=\frac{1}{2}{\rm Tr}[(\bar\bC_0^{-1} - \bar\bC_1^{-1})\brho]+\text{terms indep of }\bdt\,,
\label{e:LR_quad}
\end{equation}
where $\bar\bC_0^{-1}$ and $\bar\bC_1^{-1}$ denote the inverse covariance matrices defined by the fixed parameter values $\bar\btheta_0$ and $\bar\theta_1$.

Thus, for a simple hypothesis, the log Bayes factor is linear in the correlations $\brho$ or, equivalently, \emph{quadratic} in the timing residual data $\bdt$. This quadratic form simplifies further in the limit of weak anisotropy. Indeed, for $\bar P_2^{\rm r},\bar P_2^{\rm i}\ll P_0$ we can expand $\bar\bC_1^{-1}$ to first order in $\Gamma^{\rm aniso}$, obtaining
\begin{equation}\label
{eq:w_optimal}
    \bar\bC_0^{-1}-\bar\bC_1^{-1}\simeq\frac{\bar\sigma_{\GWB}^2}{N_s}\,\bar\bC_0^{-1}\bar\Gamma^{\rm aniso}\bar\bC_0^{-1}\,.
\end{equation}
If, in addition, the GWB is weak compared to the noise, i.e., $\bar\sigma^2_{\GWB}\ll\bar\sigma_n^2$, the inverse covariance reduces to $[\bar\bC_0^{-1}]_{ab}\simeq N_s\,\delta_{ab}/\bar\sigma_n^2$, and for a single anisotropy parameter, $\bar P_2^{\rm r}=\bar P_2^{\rm i}\equiv \bar P_2$, we obtain
\begin{equation}
\label{eq:w_optimal_weak}
    [\bar\bC_0^{-1}-\bar\bC_1^{-1}]_{ab}
    \propto -\cos(\phi_a+\phi_b+\pi/4)\,.
\end{equation}
Therefore, for simple hypotheses in the weak-signal limit, the optimal detection statistic is quadratic in the data, and its weights depend on the pulsar pair only through the combination $\phi_a+\phi_b$. All the dependence on the noise and GWB parameters is confined to an (unimportant) overall multiplicative factor, which leaves the detection statistic performance unchanged.

\subsection{Sky-map-based detection statistics}

\noindent Most frequentist statistics for anisotropy detection are constructed from an estimator of the GWB sky map (see, for example,~\cite{Taylor:2013esa,Mingarelli:2013dsa,Pol:2022sjn,NANOGrav:2023tcn,Konstandin:2025ifn}). 
Specifically, given a set of cross-correlation estimators, $\brho$, an estimator of the GWB sky map is derived by maximizing a Gaussian likelihood of the form: 
\begin{equation}\label{eq:map_likelihood}
p(\brho|P)=\frac{\exp[-\frac{1}{2}(\brho-RP)^T\Sigma^{-1}(\brho-RP)]}{\sqrt{\det(2\pi\Sigma)}}\,,
\end{equation}
where $P$ is the GWB sky map, $R$ is the response function defined in Eq.~\eqref{eq:R_response}, and $\Sigma$ is the covariance matrix of the cross-correlation estimators. For the case of our toy model, this covariance matrix is given by:
\begin{equation}
    \Sigma_{ab,cd} = N_s\left(C_{ac}C_{bd} + C_{ad}C_{bc}\right)\,.
\end{equation}
This covariance matrix depends on both the noise parameters ($\sigma_n$ for our toy model) and the GWB parameters ($\sigma_{\GWB}$ and $P_2$), none of which are known a priori. The noise parameters and the spectral properties of the GWB can be estimated from single-pulsar analyses and a preliminary Bayesian analysis modeling the GWB as common uncorrelated red noise (CURN). 

The anisotropy parameters, however, are encoded in the cross-correlations between pulsars, and therefore cannot be recovered from these preliminary, faster analyses. What is typically done instead is to neglect the anisotropic contribution altogether, approximating $\bC$ by the covariance matrix of the null hypothesis:
\begin{equation}\label{eq:rho_cov_approx}
    C_{ab}\simeq\frac{1}{N_s}\left( \delta_{ab}\,\sigma^2_n+ \Gamma_{ab}^{\rm iso}\,\sigma^2_{\GWB}\right)\,.
\end{equation}
In the following, we assume that $\sigma^2_n$ and $\sigma^2_{\GWB}$ are measured perfectly, and set them to the values $\bar\sigma^2_n$ and $\bar\sigma^2_{\GWB}$ used to generate the mock data.

The maximum-likelihood estimator for $P_2$ and $P_{-2}$ can then be found analytically and they are given by 
\begin{equation}
    \hat P_{m} = M_m^{-1} X_m,
\end{equation}
where we have defined $M_m \equiv R_{ab,m}^T \Sigma^{-1}_{ab,cd} R_{cd,m}$, and $X_m \equiv R_{ab,m}^T \Sigma^{-1}_{ab,cd} \rho_{cd}$, 
where repeated $ab$ and $cd$ pulsar pair indices are summed over. The estimators of the $P_{\pm2}$ coefficients can then be used to construct the detection statistic
\begin{equation}
    \Lambda_{\rm map} = |\hat P_2|^2 + |\hat P_{-2}|^2\,.
\end{equation}

\subsection{Neural network detection statistics}
\label{sec:NN-det-stat}
\noindent Finally, we can sidestep the construction of an explicit statistic and train a neural network to classify data as drawn from the isotropic or anisotropic hypothesis. Labeled training sets are generated by forward modeling: model parameters are drawn from the priors $\pi(\btheta_0|H_0)$ and $\pi(\btheta_1|H_1)$, and mock data are then generated from $p(\bdt|\btheta_0,H_0)$ and $p(\bdt|\btheta_1,H_1)$. As we show below, minimizing the binary cross-entropy drives the network output toward the log Bayes factor, so that a sufficiently flexible classifier recovers the optimal statistic without evaluating a single evidence integral
(see~\cite{Cranmer:2015bka,Hermans:2019oj,Jeffrey:2023stk} for a comprehensive discussion of the connection between binary classification and Bayesian model selection).

Specifically, consider a neural network $f_\lambda$ that takes as input the timing residuals, or some quantity derived from them, and returns a single number in the interval $(0,1)$, to be interpreted as the probability that the data were drawn under $H_1$. In this section we take the input to be the correlations $\rho_{ab}$; networks trained directly on the timing residuals are discussed in Appendix~\ref{sec:app_linear_quad}. Denoting collectively by $\lambda$ the trainable parameters of the network (its weights and biases), the output can be written as
\begin{equation}
    f_\lambda(\rho) \equiv \sigma\big(z_\lambda(\rho)\big)\,,
    \qquad
    \sigma(z) \equiv \frac{1}{1+e^{-z}}\,,
    \label{e:nn_output}
\end{equation}
where $z_\lambda(\rho)$ is the pre-activation output of the final neuron and
$\sigma$ is the sigmoid activation function.
For later reference, we note that
\begin{equation}
    \sigma^{-1}(u) \equiv
    \ln\left(\frac{u}{1-u}\right)
    \label{e:sig_inv}
\end{equation}
is the inverse of the sigmoid function, satisfying $\sigma^{-1}(\sigma(z))=z$.

The trainable parameters are determined by minimizing the binary cross-entropy
(BCE) functional
\begin{equation}
    \begin{aligned}
        {\rm BCE}[f_\lambda]
        &\equiv - \int\!\D \rho\!\int\!\D y\> p(\rho,y)\big[y\,\ln f_\lambda(\rho)\\
        &\hspace{.66in}+ (1-y)\,\ln(1-f_\lambda(\rho))\big]\,,
    \end{aligned}
    \label{e:BCE}
\end{equation}
where the integration is over labeled training data $(\rho, y)$, with $y=0$
(or $1$) if $\rho$ is drawn from the model evidence $p(\rho|H_0)$
(or $p(\rho|H_1)$).
The joint distribution appearing in Eq.~\eqref{e:BCE} is given by
\begin{equation}
    p(\rho,y) \equiv \frac{1}{2}\left[p(\rho|H_0)\,\delta(y)
    + p(\rho|H_1)\, \delta(1-y)\right]\,,
\end{equation}
where we have assumed equal prior odds for the two hypotheses.
Performing the integral over $y$ then yields
\begin{equation}
    \begin{aligned}
        {\rm BCE}[f_\lambda]&= -\int \D \rho\> \big[p(\rho|H_1)\,\ln f_\lambda(\rho) \\
        &\hspace{0.66in}+ p(\rho|H_0)\,\ln(1-f_\lambda(\rho))\big]\,,
        \label{e:BCE_2}
    \end{aligned}
\end{equation}
where we have dropped the overall factor of $1/2$, which does not affect the
minimization.

By minimizing this functional with respect to $f_\lambda(\rho)$, we find that the minimum is obtained when the network output satisfies
\begin{equation}
    \frac{f_\star(\rho)}{1-f_\star(\rho)} =\frac{p(\rho|H_1)}{p(\rho|H_0)}
    \equiv{\mathcal B}_{10}(\rho)\,,
\end{equation}
or, using Eq.~\eqref{e:sig_inv},
\begin{equation}
    z_\star(\rho) = \ln{\mathcal B}_{10}(\rho)\,.
    \label{e:z_opt}
\end{equation}
This theoretical result is independent of the explicit form of the network architecture of $z_\lambda$ (provided it is suitably expressive~\cite{Hornik:1989aaa}), training against the BCE drives the network toward the log Bayes factor.
However, it is important to stress that what any given architecture actually achieves is the closest approximation to the log Bayes factor within the family of functions $z_\lambda$ can represent.
Therefore, the BCE minimization effectively projects the log Bayes factor onto that family, and the performance of the resulting statistic is controlled entirely by how well the family covers it.

The simplest such family is a single linear layer, for which
\begin{equation}
    z_\lambda(\rho) = \sum_{a,b} w_{ab}\,\rho_{ab} + b
    = {\rm Tr}(w\rho) + b\,,
    \label{e:linear_nn}
\end{equation}
with $\lambda = \{w_{ab}, b\}$.
The output is linear in $\rho_{ab}$, and therefore quadratic in the timing residuals, cf.~Eq~\eqref{e:rho_ab}. By Eq.~\eqref{e:z_opt}, minimizing the BCE over this family returns the {\it best quadratic approximation} to the log Bayes factor (see also Appendix~\ref{sec:app_linear_quad}).
In what follows we refer to this architecture as the \emph{linear} network.

The second architecture we consider is a multilayer perceptron (MLP), in which
the correlations are passed through a sequence of $L$ hidden layers before
reaching the output neuron.  Specifically, the network is defined by
\begin{equation}
    \begin{aligned}
        h^{(0)} &= \rho\,,
        \\
        h^{(l)} &= \varphi\big(w^{(l)} h^{(l-1)} + \beta^{(l)}\big)\,,
        \qquad l = 1,\dots,L\,,
        \\
        z_\lambda(\rho) &= w^{(L+1)} h^{(L)} + \beta^{(L+1)}\,.
    \end{aligned}
    \label{e:mlp}
\end{equation}
Here $h^{(l)} \in \mathbb{R}^{n_l}$ is the output of the $l$-th hidden layer, $w^{(l)} \in \mathbb{R}^{n_l \times n_{l-1}}$ and $\beta^{(l)} \in \mathbb{R}^{n_l}$ are the corresponding weights and biases, and $\varphi(u) \equiv \max(0,u)$ is the ReLU activation, applied component-wise.  The widths $n_1,\dots,n_L$ of the hidden layers are hyperparameters fixed before training, while the output layer has $n_{L+1}=1$ so that $z_\lambda$ is a single number.  The trainable parameters of this network are therefore given by $\lambda = \{w^{(l)},\beta^{(l)}\}_{l=1}^{L+1}$.

Setting $L=0$ removes the hidden layers altogether, making the network collapse to a linear network equivalent to Eq.~\eqref{e:linear_nn}. However, because $\varphi$ is non-linear, $z_\lambda$ is no longer a quadratic function of the timing residuals for $L \geq 1$ and, with sufficient width, can approximate any continuous function of the correlations and find the BCE minimum defined in Eq.~\eqref{e:z_opt}.

%%%%%%%%%%%%%%%%%%%%%%%%%%%%
\section{Results}
\label{sec:results}
\begin{figure*}[ht]
\centering
\includegraphics[width=\textwidth]{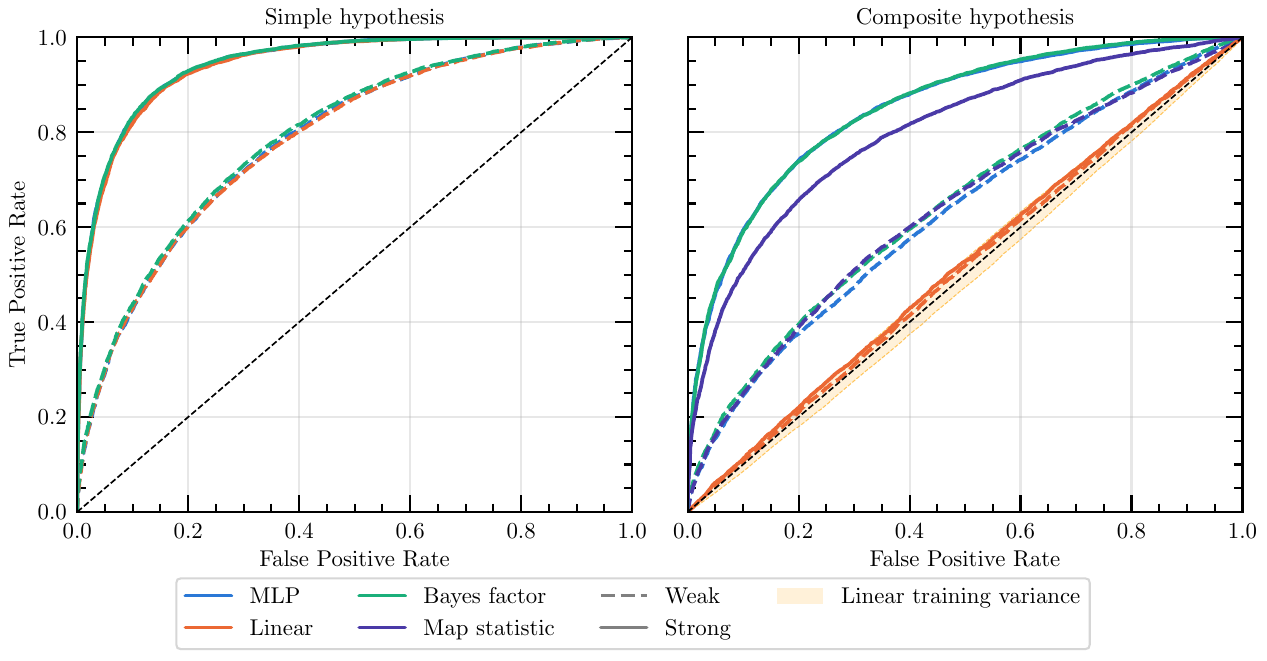}
\caption{ROC curves for the detection statistics defined in Sec.~\ref{sec:det_stats}. In the left (right) panel, we report detection statistic performances for a simple (composite) hypothesis test, both in the weak (dashed lines) and strong signal regime (solid lines) -- the specific parameter values used in these two regimes are reported at the beginning of Sec.~\ref{sec:results}. Notice that for the simple hypothesis test we do not report the performance of the map-based statistic, as the latter is constructed to estimate the anisotropy parameters from the data and therefore has no analog when those parameters are assumed known. In the right panel, we show an error band (estimated by running 20 independent realizations of the training process) associated with the training variance for the linear network.}
\label{fig:roc}
\end{figure*}
\begin{figure*}[th]
\centering
\includegraphics[width=\textwidth]{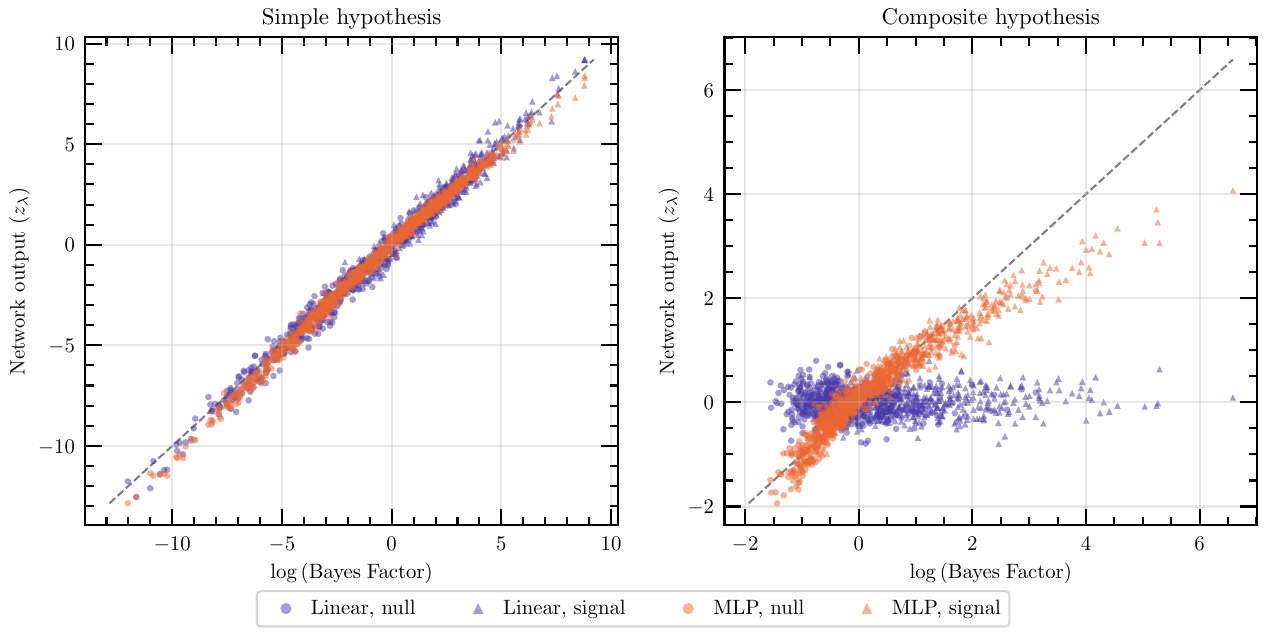}
\caption{The left (right) panel reports the comparison between the log Bayes factor and the pre-activation output of the neural network classifier, $z_\lambda$, for simple (composite) hypotheses. The $y$-coordinate of the violet (orange) markers is obtained using the linear (non-linear) network; circles (triangles) are for realizations of the null (signal) hypothesis. The dashed line marks $z_\lambda = \ln\mathcal{B}_{10}$.}
\label{fig:bf_net}
\end{figure*}
\begin{figure}[th]
\centering
\includegraphics[width=\columnwidth]{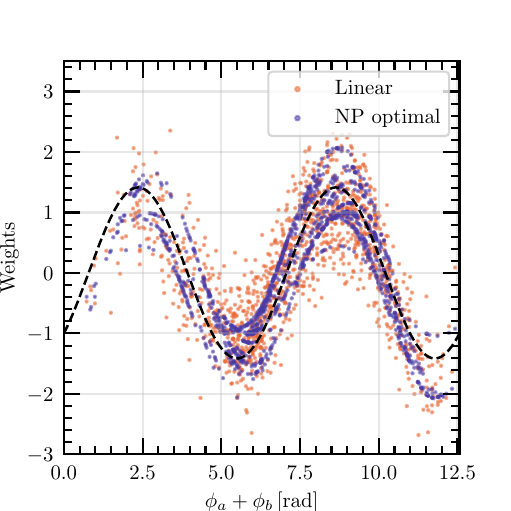}
\caption{Weights learned by the linear classifier in the weak-signal limit of a simple hypothesis test (orange) compared with the optimal weights (purple) defined in~Eq.~\eqref{eq:w_optimal}. 
The black dashed line shows the weak-signal limit of these optimal weights, cf.~Eq.~\eqref{eq:w_optimal_weak}.}
\label{fig:weights_comp}
\end{figure}
\noindent In this section, we compare the performance of the detection statistics introduced in Sec.~\ref{sec:det_stats}: the Bayes factor, the map-based statistic, and the linear and non-linear neural-network classifiers. We first consider the case of simple hypotheses, in which the noise and GWB parameters are held fixed, and then turn to composite hypotheses, in which the signal parameters are unknown and must be marginalized over.

Throughout this section, we work with an array of $N_p = 50$ pulsars, with angular positions $\phi_a$ drawn uniformly on the circle, and fix the number of measured frequencies to $N_f = 10$ and $N_s = 2 N_f = 20$. The input to both networks is the vector of $N_p(N_p+1)/2 = 1275$ independent correlations. For the non-linear network in Eq.~\eqref{e:mlp}, we use $L = 1$ hidden layer with width $n_1 = 50$. Both architectures are trained on $5\times10^4$ simulations\footnote{For the composite hypothesis (which constitutes a harder learning task) analyzed with a non-linear network, these are resampled at every epoch.} per hypothesis with a batch size of $10000$. We train the linear network with L-BFGS~\cite{LBFGS} until convergence, while we train the non-linear network with the AdamW optimizer~\cite{AdamW} and weight decay $10^{-2}$. In all cases (simple/composite and weak/strong), we use a learning rate of $10^{-3}$. To prevent overfitting, we train for up to $100$ ($300$) epochs and set an early stopping criterion after $10$ ($20$) epochs without improvement in the validation loss for the simple (composite) hypothesis.

\subsection{Simple hypotheses}
\noindent We start by considering an idealized scenario in which we test two simple hypotheses. That is, both $H_0$ and $H_1$ are fully specified, with $\bar\sigma_n^2 = 1.0$, $\bar\sigma_{\GWB}^2 = 1.0\;(0.09)$ for the strong (weak) signal case, and $\bar P_2^{\rm r} = \bar P_2^{\rm i} = 1/(4\pi\sqrt2)$ for the signal hypothesis. These values are used both to generate the mock data and to construct the detection statistics: the priors in Eq.~\eqref{e:p0} collapse to delta functions, so that the evidences reduce to likelihoods evaluated at the true parameters and every training set is drawn from a single point in parameter space rather than from a distribution over it. In this case the Bayes factor reduces to a linear function of the correlations, cf.~Eq.~\eqref{e:LR_quad}, which both the linear and the non-linear classifier should be able to approximate. 

The performance of the different detection statistics is compared using the receiver operating characteristic (ROC) curves, which quantify the trade-off between detection rate (true positive rate) and false alarm rate (false positive rate) as we vary the classification threshold. As expected, for these simple hypotheses the Bayes factor and the two classifiers perform identically in both the weak and strong signal regimes; see the left panel of Fig.~\ref{fig:roc}.\footnote{The map-based statistic is not included in this comparison, since it is designed to estimate anisotropy parameters that are here assumed known.} Indeed, in this case both the linear and non-linear classifiers can approximate the Bayes factor exactly, which is a linear function of the correlations.

In addition to comparing the ROC curves, we can also directly compare the output of the classifiers with the Bayes factor. This comparison is shown in Fig.~\ref{fig:bf_net} for a set of 1000 realizations of the null and signal hypothesis. From this figure, we can see that for simple hypotheses the output of both the linear and non-linear classifiers is linearly related to the log Bayes factor. 

Finally, in the weak-signal limit we expect the weights of the linear network to depend on the pulsar pair only through the combination $\phi_a+\phi_b$, reducing to the simple form of Eq.~\eqref{eq:w_optimal_weak}. In Fig.~\ref{fig:weights_comp} we compare the actual network weights learned by the linear classifier with this analytic prediction. We find that, up to an overall normalization, the two agree: the learned weights trace the expected $\cos(\phi_a+\phi_b+\pi/4)$ profile, with a scatter consistent with the finite size of the training set.

All these results establish that both classifiers (linear and non-linear networks) recover the optimal statistic for simple hypotheses, and in the linear case they do so by learning the Neyman-Pearson weights themselves. 

\subsection{Composite hypotheses}
\noindent We now turn to a more realistic case---closer to a real anisotropy search---in which the parameters entering the two hypotheses are not known a priori. Specifically, we keep $\sigma_n$ and $\sigma_{\GWB}$ fixed to the value used in the simple hypothesis case, but treat the anisotropic parameters $P_2^{\rm r}$ and $ P_2^{\rm i}$ as unknown. When generating mock data for the signal hypothesis, we draw the anisotropy parameters from a uniform prior over the disk $\sqrt{(P_2^{\rm r})^2 + (P_2^{\rm i})^2} \le P_2^{\max}$, with $P_2^{\max} = 1/(4\pi)$, the largest anisotropy for which the GWB power remains non-negative over the whole sky. The same prior is used to compute the Bayes factor and generate the training data for the neural network classifiers. 
Thus, the log Bayes factor is no longer the likelihood ratio evaluated at fixed parameters, and, as anticipated in Sec.~\ref{sec:det_stats}, it is no longer a linear function of the correlations. We therefore expect the linear classifier to underperform the Bayes factor, while the (non-linear) MLP should be able to match it.

These expectations are confirmed by the right panel of Fig.~\ref{fig:roc}: the non-linear classifier tracks the Bayes factor across the full range of false alarm probabilities, while the linear one performs no better than chance. This is a direct consequence of the zero-mean prior on the anisotropy parameters. Averaged over this prior, the expected correlations are identical under the two hypotheses, so no statistic linear in $\rho_{ab}$ can distinguish them, and the BCE is minimized at $w=0$ (see App.~\ref{sec:app_linear_quad}). The map-based statistic, which we now include in the comparison, performs better than the linear classifier but still falls short of the optimum. Its deficit has a different origin: it stems from treating the cross-correlations as Gaussian variables and from the approximation in Eq.~\eqref{eq:rho_cov_approx}, where the contribution of the anisotropic component to the covariance matrix is neglected~\cite{Lemke:2026gxj}. Notice that the gap between the non-linear classifier and the map-based statistic is smaller here than in Ref.~\cite{Lemke:2026gxj}, which used a single frequency bin. Our cross-correlations sum over $N_s = 20$ bins, cf.~Eq.~\eqref{e:rho_ab}, which drives their distribution toward a Gaussian and brings the map-based statistic, built on that assumption, closer to optimality.

Finally, as for simple hypotheses, we can compare the output of the classifiers directly with the log Bayes factor. Consistent with the discussion above, the right panel of Fig.~\ref{fig:bf_net} shows that the output of the linear network is uncorrelated with the Bayes factor, scattering around zero.
The non-linear classifier, on the other hand, tracks the log Bayes factor up to values of $\log \mathcal{B}_{10}\sim 2$ where it starts to become under-confident. We believe that this stems from the likelihood becoming more and more peaked for larger BF values, and therefore requiring a larger number of training samples (and/or a more expressive network, i.e., more layers/parameters) for the network to approximate it properly. 

\section{Conclusions}
\label{sec:conclusions}
\noindent Simulation-based inference (SBI) is increasingly used in GW data analysis (see, e.g., Refs.~\cite{Shih:2023jme, Lai:2025xov, Laal:2024trp, Lemke:2026gxj, Dandapat:2026kfd} for applications to PTA data analysis). Reference~\cite{Lemke:2026gxj} shows SBI methods can outperform classical frequentist methods while remaining far faster than a full Bayesian analysis for GWB anisotropy searches. Given the somewhat black-box nature of these methods, and the significance of the claims that would follow from a detection, here we aim to further demonstrate why they work and where their advantage over frequentist statistics comes from. In this work, we introduced a toy PTA model in which full Bayesian analyses can be performed quickly, making it possible both to benchmark SBI against them and to inspect what the networks have learned.

Specifically, we considered anisotropy detection and compared the performance of the Bayes factor, the sky-map-based statistic commonly used in real searches, and two neural classifiers: a linear one, which is restricted to quadratic functions of the timing residuals, and a multilayer perceptron with more functional freedom. We find that during training the classifiers are (as expected) driven toward the closest approximation to the log Bayes factor within the family of functions their architecture can represent. For simple hypotheses, the Bayes factor is itself quadratic in the timing residuals (see Sec.~\ref{sec:BF_det_stats}), so both classifiers reach it and perform identically. Once we must marginalize over the anisotropy parameters, this is no longer the case, and only the MLP can efficiently approximate the log Bayes factor. For the linear network, we can go further than comparing performance. Since the learned statistic is fully specified by its weights, we can compare them directly with the analytic Neyman-Pearson prediction. For this case, we find that the two agree up to an overall normalization, as discussed in Sec.~\ref{sec:results}.

To conclude, in this pedagogical work we have used a simple PTA toy model as a case study to understand what neural classifiers learn when used as a detection statistic for GWB anisotropies, and we found that training such a classifier amounts to constructing a frequentist detection statistic that, given sufficient network flexibility, reaches the sensitivity of a full Bayesian analysis at a fraction of its cost. 
We hope that this work will be useful to those approaching these new techniques for the first time, offering some intuition for what SBI is doing and how it does it.\\

\noindent\emph{Data availability --} The code and data needed to reproduce the results in this paper are available at \href{https://github.com/Mauropieroni/PTAs_in_flatland}{this link}. 
\bigskip 
\acknowledgments
\noindent We thank A.M. Lemke and T. Konstandin for several insightful discussions about related topics. AM acknowledges support from a Royal Society University Research Fellowship
(URF-R1-251896). JDR acknowledges financial support from NSF Physics Frontier Center Award PFC-2020265 and start-up funds from the University of Texas Rio Grande Valley. The work of MP is supported by the Comunidad de Madrid under the Programa de Atracción de Talento Investigador with number 2024-T1TEC-31343. MP acknowledges the hospitality of Imperial College London, which provided office space during parts of this project. JA is supported by a fellowship from the Kavli Foundation.
AM, MP, and JDR also acknowledge the hospitality of the Galileo Galilei Institute during the workshop  “New Frontiers in Gravitational-Wave Physics”, where part of this work was carried out.
\bigskip

\appendix

\section{When can a linear neural network learn from training data?}
\label{sec:app_linear_quad}

\noindent In this appendix, we discuss when a single-layer (linear) neural network can provide a (quadratic) approximation to the log Bayes factor.
We will consider using either the individual timing residuals $\bdt_a$ or the correlations $\rho_{ab}$ as the input data to the linear neural network. 
In particular, we show that a linear neural network learns nothing from the timing residual data.
This is because a linear network  is only capable of building statistics \emph{linear} in the input data, while the information regarding the GWB is encoded in the \emph{quadratic} combinations (i.e., the correlations) of the timing residual data.

In general, a linear neural network can only perform an affine transformation of the data. 
The most general form of such a transformation is given by 
\be
z_\lambda(x) = w\cdot x + b\,,
\label{e:z(x)}
\ee
with $\lambda=\{w,b\}$ denoting the learnable vector of weights $w$ and bias $b$, respectively. 
The weights and bias are typically learned by minimizing a cost function, such as the binary cross-entropy (BCE) given in~Eq.~\eqref{e:BCE_2}, which we  rewrite as 
\begin{equation}
    \begin{aligned}
        {\rm BCE}[f_\lambda]&= -\int \D x\> \big[p(x|H_1)\,\ln f_\lambda(x) \\
        &\hspace{0.66in}+ p(x|H_0)\,\ln(1-f_\lambda(x))\big]\,,
        \label{e:BCE_x}
    \end{aligned}
\end{equation}
with the input data denoted by $x$, which could be either the correlations $\rho_{ab}$ or the timing residuals $\bdt_a$.
Similar to before, we define
$f_\lambda(x)\equiv \sigma(z_\lambda(x))$, where $\sigma(z)$ denotes the sigmoid function, see Eq.~\eqref{e:nn_output}.

Since $-\ln \sigma(z_\lambda(x))$ is convex, the BCE cost function is also convex, and therefore it has a unique minimum. 
By varying the BCE cost function with respect to the weights and bias, we find:
\begin{equation}
\begin{aligned}
\delta{\rm BCE}[f_\lambda]&= 
-\int \D x\> \big[p(x|H_1)(1- \sigma(z_\lambda(x))) \\
        &\hspace{0in}- p(x|H_0)\,\sigma(z_\lambda(x))\big](\delta w\cdot x +\delta b)\,,
        \label{e:deltaBCE}
\end{aligned}
\end{equation}
where we used the result that the derivative of the sigmoid function satisfies $\sigma'(z)= \sigma(z)(1-\sigma(z))$.
Evaluating the variation at $w = 0$ and $b = 0$, we find:
\be
\delta{\rm BCE}[f_\lambda]\big|_{w=0,b=0} =
-\frac{1}{2}\left(\langle x\rangle_{\bar 1}-\langle x\rangle_{\bar 0}\right)\cdot \delta w\,,
\label{e:deltaBCE_simple}
\ee
where $x$ is averaged over the model evidences:
\be
\begin{aligned}
&\langle x\rangle_{\bar 1}\equiv\int\D\theta_1\>\pi(\theta_1|H_1)\int\D x\>
x\,p(x|\theta_1,H_1)\,,\\
&\langle x\rangle_{\bar 0}\equiv\int\D\theta_0\>\pi(\theta_0|H_0)\int\D x\>
x\,p(x|\theta_0,H_0)\,.
\end{aligned}
\ee
Note that Eq.~\eqref{e:deltaBCE_simple} does not depend on $\delta b$.

\begin{figure*}[ht]
\centering
\includegraphics[width=\textwidth]{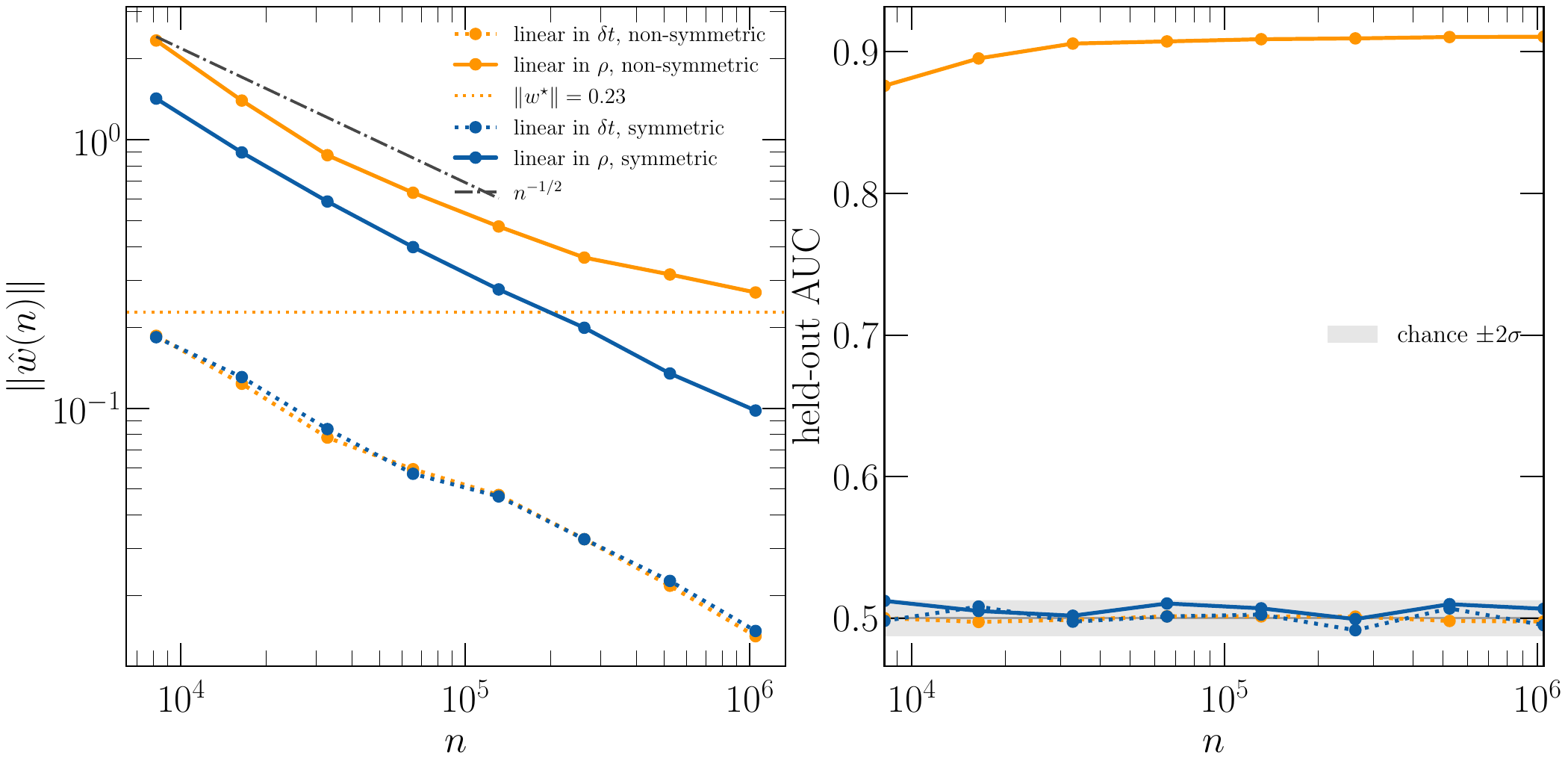}
\caption{ 
Left: the average weights $\hat{w}(n)$ of a linear network go to zero as the number of training samples $n$ increases when using either the timing residuals or the correlations (with zero-mean priors) for the input data to the network. On the other hand, the weights of a linear network trained on quadratic data only go to zero when the prior is zero-mean.
Right: the AUC (area under the ROC curve) is 50\% for those cases where the weights tend to zero. Only the linear network trained on quadratic data with a non-zero mean prior returns a non-trivial detection statistic.
\label{fig:scaling}} 
\end{figure*}

If we take the input data $x$ to be the set of timing residuals $\delta t_a$, then $\langle \bdt_a\rangle_{\bar 1}=\langle \bdt_a\rangle_{\bar 0}=0$ for any choice of the prior probability distributions since the timing residuals are zero-mean Gaussian random variables.
Thus, 
$\delta{\rm BCE}[f_\lambda]\big|_{w=0,b=0} =0$ for this case, and the linear neural network learns nothing from the data as shown in Fig.~\ref{fig:scaling}.
The global minimum of the BCE loss function is at $w=0$, $b=0$.

If, instead, we take $x$ to be the set of correlations $\rho_{ab}$, then one can show that
\be
\langle \rho_{ab}\rangle_{\bar 1}-\langle \rho_{ab}\rangle_{\bar 0}
\propto\int\D\theta_1\>\pi(\theta_1|H_1)\sigma^2_{\GWB}\,\Gamma_{ab}^{\rm aniso}\,,
\ee
where $\theta_1$ is given in Eq.~\eqref{e:params1} and $\Gamma_{ab}^{\rm aniso}=\Gamma_{ab}^{\rm aniso}(P_2^{\rm r}, P_2^{\rm i})$, see Eq.~\eqref{e:Gamma_aniso}.
If we have zero-mean priors on $P_2^{\rm r}$ and $P_2^{\rm i}$ then $\delta{\rm BCE}[f_\lambda]\big|_{w=0,b=0} =0$
for this case as well, and the BCE loss function again has its minimum at $w=0$, $b=0$, see Fig.~\ref{fig:scaling}.
It is only when we have nonzero-mean priors on $P_2^{\rm r}$ and $P_2^{\rm i}$ that we obtain a non-zero variation:
\be
\begin{aligned}
&\delta{\rm BCE}[f_\lambda]\big|_{w=0,b=0}\\
&\hspace{0.15in}\propto\sum_{a,b}\left(\bar P_2^{\rm r}\cos(\phi_a+\phi_b) - \bar P_2^{\rm i}\sin(\phi_a+\phi_b)\right)\delta w_{ab}\,.
\end{aligned}
\ee
For this case, the linear neural network is able to learn non-trivial weights under training, and we obtain a quadratic approximation to the log Bayes factor as desired.

\bibliographystyle{utphys}
\bibliography{ref}
\end{document}